\documentclass[aps,prl,twocolumn,superscriptaddress,raggedbottom,showpacs,floatfix]{revtex4-1}
\usepackage{amsmath,amsfonts,amssymb,amsthm,graphics,graphicx,latexsym,color}
\usepackage[next]{inputenc}
\usepackage[dvips]{epsfig}
\usepackage[colorlinks=true,citecolor=blue,linkcolor=blue]{hyperref}
\usepackage{bbm,bm, bbold}
\usepackage{booktabs}
\usepackage{multirow}
\usepackage{hhline}
\usepackage{comment}
\usepackage{float}

\usepackage{epstopdf}

\begin{document} 
\renewcommand{\vec}{\mathbf}
\renewcommand{\Re}{\mathop{\mathrm{Re}}\nolimits}
\renewcommand{\Im}{\mathop{\mathrm{Im}}\nolimits}

\title{Helical Network Model for Twisted Bilayer Graphene}
\author{Dmitry K. Efimkin}
\affiliation{The Center for Complex Quantum Systems, The University of Texas at Austin, Austin, Texas 78712-1192, USA}

\author{Allan H. MacDonald}
\affiliation{The Center for Complex Quantum Systems, The University of Texas at Austin, Austin, Texas 78712-1192, USA}

\begin{abstract}
In the presence of a finite interlayer displacement field bilayer graphene has an 
energy gap that is dependent on stacking and largest for the stable AB and BA stacking arrangements.
When the relative orientations between layers are twisted through a small angle to form a moir$\mathrm{\acute{e}}$ pattern, the local
stacking arrangement changes slowly.  We show that for non-zero displacement fields
the low-energy physics of twisted bilayers 
is captured by a phenomenological helical network model that 
describes electrons localized on domain walls separating regions with approximate AB and BA stacking. The network band structure is gapless and has of a series of two-dimensional bands 
with Dirac band-touching points
and a density-of-states that is periodic in energy with one zero and one divergence per period.
\end{abstract}
\maketitle

\noindent
\emph{Introduction}---
The electronic structure of bilayer graphene is
sensitive to strain, interlayer potential differences, and the stacking arrangement between 
layers~\cite{BGReview2,BGReview1}. 
For the energetically favored Bernal stacking configurations, either $\mathrm{AB}$ or $\mathrm{BA}$,  
Bloch states have $2 \pi$ Berry phases, quadratic band-touchng, and a gap that opens
when a displacement fields is applied by external gates.  
The gapped state is characterized by nontrivial valley-dependent Chern numbers 
and supports topological confinement of electrons on domain walls that separate 
regions with opposite signs of displacement field ~\cite{BGdomainMartin,BGdomainNunez,BGdomainCosma,BGdomainCosta} or different stacking arrangements~\cite{MacDonaldZhang,BGdomainVaezi,BGdomainKoshino}.
The presence of confined electronic states,
which occur in helical pairs with opposite propagation directions in opposite valleys,
has~\cite{BGdomainExp1,BGdomainExp2,BGdomainExp3} been 
confirmed experimentally. Control of these domain walls and of their intersections has attracted 
attention recently ~\cite{MacDonaldQiao1,MacDonaldQiao2,BGdomainPan,BGdomainWright,BGdomainRen,BGdomainMosallanejad} 
because of its potential relevance for valleytronics~\cite{2DReview}.

Whereas an engineering of a network of helical states with tunable geometry is a challenging problem, the triangular one has been recently observed~\cite{2018arXiv180202999H} with help of scanning tunneling spectroscopy (STM) in misoriented graphene bilayers~\cite{MacDonaldBistritzer,MacDonaldJung, GBTwist2,GBTwist3,GBTwist4,GBTwist5,GBTwist6,GBTwist7,GBTwist8,GBTwist9,BGdomainXintao,BGtwistExp1,BGtwistExp2,BGtwistExp3,BGtwistExp4,BGtwistExp5,BGtwistExp6,BGtwistExp7,BGtwistExp8,BGtwistExp9}. In the presence of a twist local stacking arrangement changes slowly in space in a periodic moir$\mathrm{\acute{e}}$ pattern in which regions with approximate $\mathrm{AB}$ and $\mathrm{BA}$ stacking are separated by domain walls with helical states	. The measured local density of states at a domain wall is strongly energy dependent with a single peak within the gap, that demonstrates the importance of an interference between helical states propagating along network. Because the moir$\mathrm{\acute{e}}$ pattern is well developed only when its period greatly exceeds graphene's lattice constant, theories of its electronic structure~\cite{BGHelicalNetwork,GBTwistBias1}
often employ complicated multi-scale approaches to advantage.
  
In this Letter, we derive a phenomenological helical network model for 
the electronic structure of gated bilayer graphene moir$\mathrm{\acute{e}}$s valid in the energy range 
below the $\mathrm{AB}$ and $\mathrm{BA}$ gaps where only topologically
confined domain wall states are present.    
The model is related to Chalker-Coddington type models~\cite{Network1,Network2,Network3} 
introduced in theories of the quantum Hall effect. 
The spectrum of the network model consists of a set of minibands connected by 
Dirac band touching points, which repeats and is gapless. 
A single period of the model's band structure is illustrated in Fig.~1.

\begin{figure}[t]
	\label{Fig1}
	\vspace{-2 pt}
	\includegraphics[width=8.5 cm]{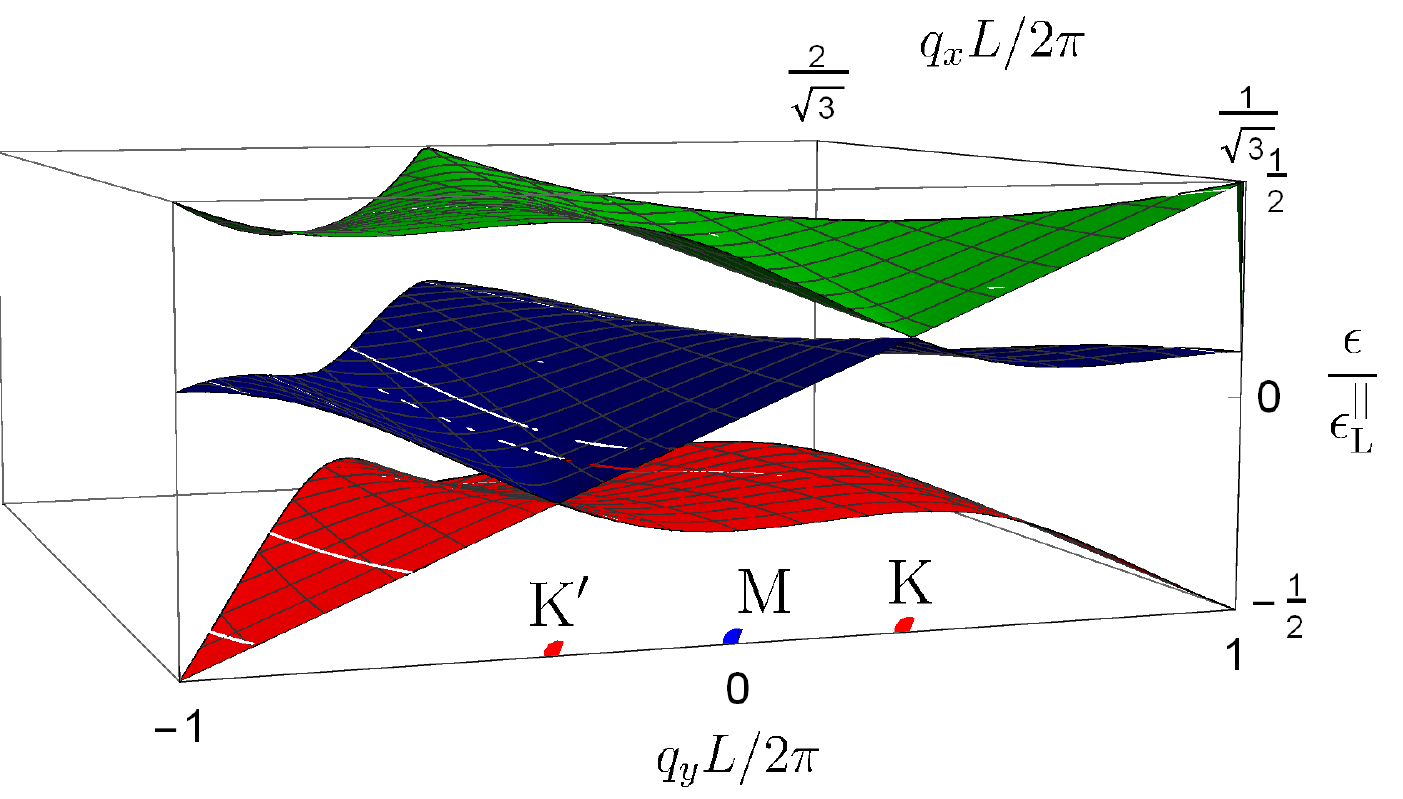}
	\vspace{-2 pt}
	\caption{Helical model band structure over half of the rhombic Brillouin zone (BZ) defined in Fig.~3-(c).
	The bands in the other half of the BZ can be obtained by the reflection.  The model's band 
	 energies $\epsilon^{n0}_\vec{q}$ are given by Eq.~(\ref{BandStructure}) and depend on a single 
	 controlling parameter $\alpha$ which was set to $\alpha=1.1$ in this illustration.
	 The bands touch at Dirac points located at high symmetry $\mathrm{K}$, $\mathrm{K}'$ and $\mathrm{\Gamma}$ points. }
\end{figure}

\noindent
\emph{Moir$\mathrm{\acute{e}}$ pattern and helical states}---
To describe the electronic structure of gated bilayer graphene 
with a small twist angle $\theta\lesssim 1^{\circ}$~\cite{NoteShift} between layers, 
we start from the continuum model Hamiltonian derived in Ref.~\cite{MacDonaldBistritzer},
which is valid independent of atomic scale commensurability
\begin{figure*}[t]
	\label{Fig2}
	\vspace{-2 pt}
	\includegraphics[width=16.4 cm]{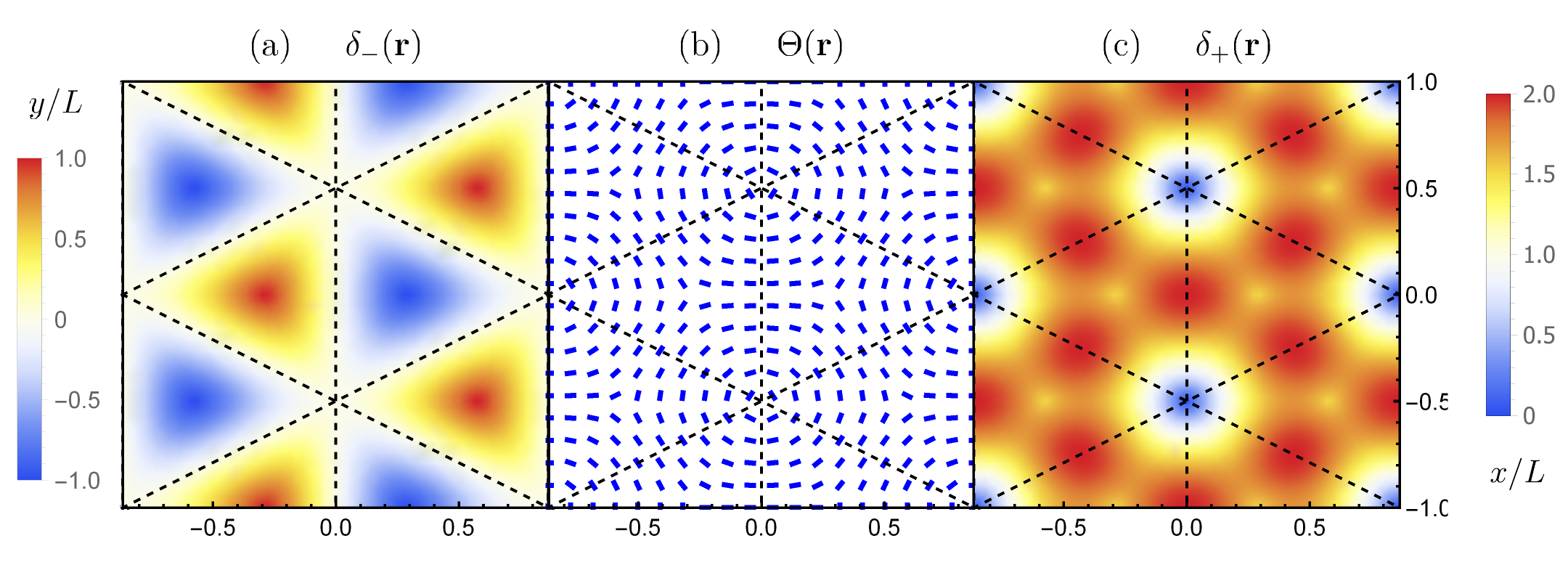}
	\vspace{-2 pt}
	\caption{Spatial distribution of the the gap parameters 
	in Eq.(\ref{Gap}): $\hbox{(a)}$ the gap minimum $\delta_-$; $\hbox{(b)}
	$ the angle $\theta(\vec{r})$ which specifies the direction in momentum space at which 
	minima are achieved;  $\hbox{(c)}$ the gap maximum  $\delta_+$.  The dashed lines highlight the 
	network of domain walls that separate regions in which the hybridization is dominated by $T_\mathrm{AB}$ 
	from regions in which it is dominated by $T_\mathrm{BA}$.
	}
\end{figure*}
\begin{equation}
\label{HamiltonianBilayer}
H_0=
\begin{pmatrix}
v \sigma_\mathrm{t} \vec{p}-u & T(\vec{r})
\\ T^+(\vec{r}) & v \sigma_\mathrm{b} \vec{p}+u
\end{pmatrix}.
\end{equation}
The Hamiltonian for a valley $K$ acts in the sublattice space
$\psi=\{\psi_\mathrm{A}^\mathrm{t},\psi_\mathrm{B}^\mathrm{t}, \psi_\mathrm{A}^\mathrm{b},\psi_{\mathrm{B}}^\mathrm{b}\}$, where $\mathrm{t}$ and $\mathrm{b}$ refer to the top and bottom layer,
$v$ is the single-layer Dirac velocity; $\sigma_\mathrm{t(b)}$ is the vector of Pauli matrices rotated by the 
angle $\pm \theta/2$ in top and bottom layers, and $2 u$ is the potential difference between layers 
produced by the gates.  The spectrum is valley and spin independent, while electronic states in two valleys $K$ and $K'$ transform to each other by the time-reversal transformation. The inter-layer hopping operator  is given by
\begin{equation}
\label{HamiltonianHybrid}
T(\vec{r})=\frac{w}{3}\sum_{i=1}^3 e^{- i \vec{k}_i \vec{r}} T_i,
\end{equation}
where $w$ is a hybridization energy scale.  The vectors $\vec{k}_1=-k_\mathrm{\theta}\vec{e}_y$, $\vec{k}_{2,3}=k_\mathrm{\theta} (\pm\sqrt{3}\vec{e}_\mathrm{x}+\vec{e}_\mathrm{y})/2$ all have magnitude 
equal to the twist-induced separation between the Dirac points of the two-layers,
$k_\mathrm{\theta}=2 k_\mathrm{D} \sin(\theta/2)$ where 
$k_\mathrm{D}=4\pi/3 a_0$ is the magnitude of the Brillouin-zone corner vector of a single layer and $a_0$ is the corresponding Bravais period.  The matrices $T_i$ are given by      
\begin{equation*}
T_1=\begin{pmatrix}
1 & 1\\
1 & 1
\end{pmatrix}, \;\;
T_2=\begin{pmatrix}
e^{- i \zeta} & 1\\
e^{ i \zeta} & e^{- i \zeta}
\end{pmatrix}, \;\;
T_3=\begin{pmatrix}
e^{i\zeta} & 1\\
e^{- i \zeta} & e^{i \zeta}
\end{pmatrix},
\end{equation*}
with $\zeta=2\pi/3$. 
The inter-layer hopping operator 
in Eq.(\ref{HamiltonianHybrid}) is spatially periodic
with the period of the moir$\mathrm{\acute{e}}$ pattern
$L=a_0/(2\sin(\theta/2))$. 

The network model we derive has its widest range of applicability in the large gate voltage regime 
$\epsilon_\mathrm{L}\ll u \sim w$
where $\epsilon_\mathrm{L}=2\pi \hbar v/L$ is the energy scale of the network mini-bands,
as we explain below.  
In this limit an energy gap $\sim w$ develops around the momentum space
ring of radius $p_\mathrm{u}=u/v$ where the the conduction band of the 
low potential top layer overlaps with the valence band of the high potential bottom layer.  
At energies $\epsilon \ll w$ the bilayer spectrum can be described by the projected two-band 
Hamiltonian  
\begin{equation}
\label{HamiltonianProjected}
H=
\begin{pmatrix}
v (p-p_\mathrm{u}) & t_\mathrm{P}+t_\mathrm{S} \\ 
t_\mathrm{P}^*+t_\mathrm{S}^* & -v (p-p_\mathrm{u}) \end{pmatrix}.
\end{equation}
In Eq.~\ref{HamiltonianProjected} we have 
separated the tunneling matrix element into two parts, an anisotropic part 
with $\mathrm{p}$-wave symmetry $t_\mathrm{P}(\phi_\vec{p},\vec{r})=[T_\mathrm{BA}e^{- i \varphi_\vec{p}}-T_\mathrm{AB}e^{ i \varphi_\vec{p}}]/2$, 
where $\varphi_\vec{p}$ is the direction of a momentum $\vec{p}$, and
an  isotropic part  $t_\mathrm{S}(\vec{r})=-i T_\mathrm{AA}(\vec{r}) \sin(\theta/2)$ independent of $\varphi_\vec{p}$ that can be neglected~\cite{NoteTd} for $\theta\ll 1$. 
The resulting local spectrum $\epsilon_{\vec{p} \pm}=\pm \sqrt{(v p-u)^2+\Delta_\vec{p}^2}$ has an
anisotropic gap
\begin{equation}\label{Gap}
\Delta_\vec{p}^2=\delta^2_{-} \cos^2[\varphi_\vec{p}-\Theta]+\delta^2_{+}\sin^2[\varphi_\vec{p}-\Theta].
\end{equation}
which achieves minima $|\delta_-|=|(|T_\mathrm{AB}|-|T_\mathrm{BA}|)/2|$ at momentum orientations 
$\varphi_\mathrm{I}=\Theta$ and $\varphi_\mathrm{II}=\Theta+\pi$, 
where $\Theta(\vec{r})=(\arg[T_\mathrm{BA}]-\arg[T_\mathrm{AB}])/2$.
The gap is maximized at $\delta_+(\vec{r})=(|T_\mathrm{BA}|+|T_\mathrm{AB}|)/2$ at
the two perpendicular orientations.  

It follows from the preceding analysis that the gap in the local electronic spectrum (\ref{Gap}) closes if 
$|T_\mathrm{AB}|=|T_\mathrm{BA}|$.  This condition is satisfied along the 
domain walls specified by dashed lines in Fig.~2-$\hbox{(a)}$, where we illustrate 
the spatial pattern of $\delta_-(\vec{r})$. 
The domain walls separate regions where the inter-layer hybridization is dominated 
by the $T_\mathrm{AB}$ from regions in which it is dominated by $T_\mathrm{BA}$.
The local valley Chern number of Hamiltonian (\ref{HamiltonianProjected}) 
 \begin{equation}
C=\int \frac{d \vec{p}}{4\pi} \;\vec{d} \left[\frac{\partial \vec{d}}{\partial p_x} \times \frac{\partial \vec{d}}{\partial p_y}\right]=\frac{\delta_{-}}{|\delta_{-}|},
\end{equation}
where $\vec{d}=\vec{h}/h$ and the vector $\vec{h}$ is defined by the Pauli matrix expansion 
of Eq.~(\ref{HamiltonianProjected}), $H=(\bm{\sigma} \cdot \vec{h})$. 
The valley Chern number difference across the domain wall is $C_\mathrm{AB}-C_\mathrm{BA}=2$,
guaranteeing that two helical electronic channels are present in the gaps per valley and per spin.

In the vicinity of each domain wall the low-energy states are concentrated around the 
minima at orientations $\varphi_\mathrm{I(II)}$, which are perpendicular to the domain wall,
as illustrated in Fig.~2-$\hbox{(b)}$. 
The expansion of the Hamiltonian (\ref{HamiltonianProjected}) in the vicinity of these minima results in a pair of 
identical anisotropic Dirac cones with spatially depended mass $\delta_-(\vec{r})$:
\begin{equation}
\label{HamiltonianDirac}
H_{D}=\begin{pmatrix}
\delta_-(\vec{r}) & v p_\perp-i v_{||} p_{||} \\
v p_\perp+i v_{||} p_{||}& -\delta_-(\vec{r})  
\end{pmatrix}.
\end{equation} 
Here the velocity for momenta $p_\perp$ perpendicular 
to the domain wall is the single-layer graphene Dirac velocity $v$.
The velocity for momenta $p_{||}$ along 
the domain wall  can be approximated by its value at the domain wall center
 $v_{||}=\delta_+/p_\mathrm{u}\approx 2 w v/3 u$.
Each Dirac point carries one half of the valley Chern number $C_\mathrm{D}=\delta_-/2|\delta_-|$, and is 
responsible for a single helical state.  The Dirac mass $\delta_-(\vec{r})$ changes sign across the domain wall and
Eq.(\ref{HamiltonianDirac}) therefore has a Jackiw-Rebbi~\cite{JackiwRebbi} solution
that describes helical electronic states with 
dispersion $\epsilon_{p_{||}}=v_{||} p_{||}$, and wave function
\begin{equation}
\psi_{p_{||}}(r_\perp)= N \begin{pmatrix} 1 \\
i
\end{pmatrix}\exp\left[i \frac{p_{||} r_{||}}{\hbar} - \frac{w L}{\pi \hbar v} \sin^2\left( \frac{\pi r_\perp}{\sqrt{3} L}\right) \right],
\end{equation}
where $N$ is a normalization factor. The center of AB/BA region, where wave functions of helical states from different domain walls overlap, are distanced at length $r_\perp^0=L/2\sqrt{3}$ from them. The domain wall network is well developed if the overlap of wave functions $|\psi_{p_{||}}(r_\perp^0)|^2|/|\psi_{p_{||}}(0)|^2=\exp[-w/\epsilon_\mathrm{L}]\ll1$ is weak. Here $\epsilon_\mathrm{L}=2\pi \hbar v/L$ is the character energy scale of the moir$\mathrm{\acute{e}}$ pattern.

\begin{figure}[t]
	\label{Fig3}
	\vspace{-2 pt}
	\includegraphics[trim={2.8cm 11.5cm 8.4cm 1.0cm},clip , width=8.9 cm]{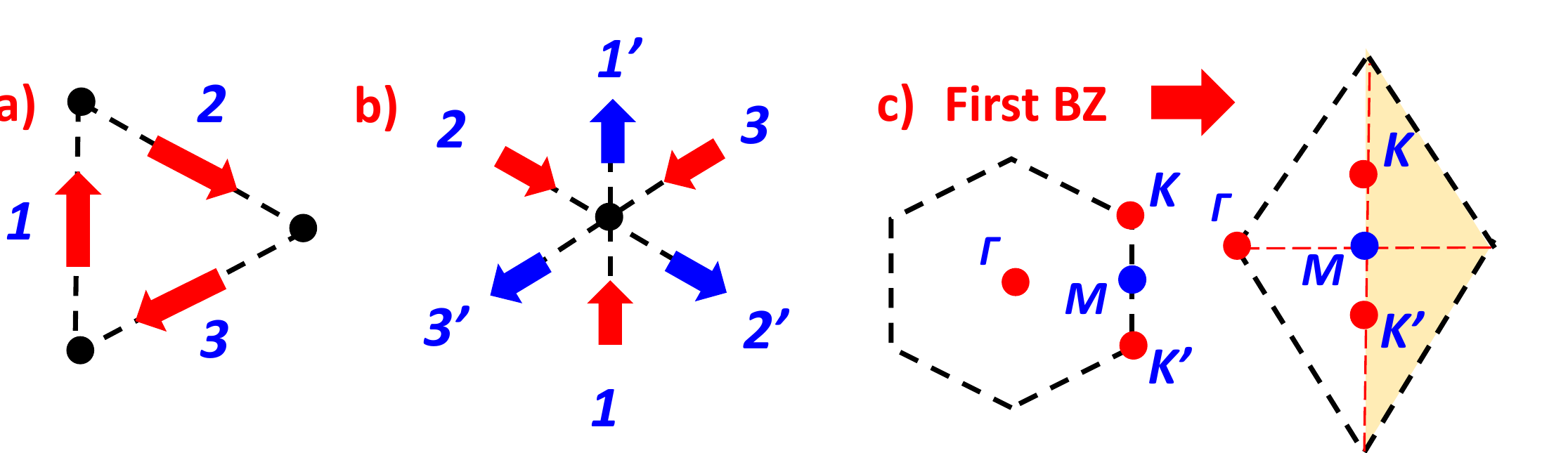}
	\vspace{-2 pt}
	\caption{$\hbox{(a)}$ Elementary cell of the network.  The wavefunction amplitudes are 
	links $1$, $2$ and $3$ are $\psi_{ij}=\left\{\psi_{ij}^1, \psi_{ij}^2, \psi_{ij}^3 \right\}$. 
	$\hbox{(b)}$ Node with three incoming and three outoing channels
	characterized by the scattering matrix $T$.  
	$\hbox{(c)}$ First Brillouin zone of the network 
	in hexagonal and rhombohedral representations.}
\end{figure}
 
These helical states are the only electronic degrees of freedom present when $|\epsilon|\ll u, w$.
Three sets of parallel domain walls with orientations differing by $120^{\circ}$ surround 
$\mathrm{AB}$ and $\mathrm{BA}$ regions and intersect at a set of points with local $\mathrm{AA}$ 
stacking. The considerations we have discussed to this point establish 
the physical picture we use to motivate our phenomenological helical network model for 
domain wall states. 

\noindent
\emph{Phenomenological network model}--- Our phenomenological helical network model consists of the
links and nodes illustrated in Fig.~3-$\hbox{(a)}$ and $\hbox{(b)}$, which connect to form the domain wall pattern. We
assume ballistic propagation along links and scattering only at nodes.
The dispersion law along links, $\epsilon=v_{||} q$, is consistent with the Jackiw-Rebbi 
confined mode solution.  For $\epsilon_\mathrm{L}\ll w\lesssim u$, the
two Dirac cones on opposite sides of the ring at $\varphi_\mathrm{I}$  
and $\varphi_\mathrm{II}$ are well separated, allowing 
scattering between them to be neglected.  This simplification 
allows us to consider a network with a single helical channel per link.

The full domain wall network can be constructed by placing the set of three 
elementary nodes on a triangular lattice with elementary lattice vectors
$\vec{l}_{1,2}=L (\pm\sqrt{3} \vec{e}_x  + \vec{e}_y)/2$. 
The wavefunction amplitudes on links $1$, $2$ and $3$ of the cell centered at 
$\vec{R}_{ij}=i \vec{l}_1 + j \vec{l}_2$ are denoted by
$\psi_{ij}=\left\{\psi_{ij}^1, \psi_{ij}^2,\psi_{ij}^3\right\}$. 
Each node has three input and three output channels and therefore 
has a $3\times3$ unitary scattering matrix $T$ whose detailed form 
depends in a complex way~\cite{NodAnglin} on the spatial profile of the domain walls intersection. We follow a simpler phenomenological approach.  By observing that the straight-forward 
scattering amplitude magnitudes
$|T_{11}|=|T_{22}|=|T_{33}|$ and the 240$^{\circ}$ deflection scattering
amplitudes $|T_{12}|=|T_{13}|=|T_{21}|=|T_{23}|=|T_{31}|=|T_{32}|$
must be equal due to symmetry, it follows that 
the unitary matrix $T$ can be parametrized by an  
angle $\alpha$ ranging between $0$ and $\alpha_\mathrm{M} = \arccos[1/3]$, 
and $6$ phases 
$\phi_\mathrm{T}, \phi_1^\mathrm{R},\phi_1^\mathrm{L},  \phi_2^\mathrm{R},\phi_2^\mathrm{L}, \phi_3$ 
ranging between $0$ and $2\pi$: 
$T=e^{i \phi_\mathrm{T}} T^\mathrm{L}_\phi \bar{T} T^\mathrm{R}_\phi$, where $\phi_\mathrm{T}$ is 
the average phase shift; $T^\mathrm{L}_\phi=\mathrm{diag}[e^{i(\phi_2^\mathrm{R} + \phi_1^\mathrm{R}+\phi_3)}, 
e^{-i\phi_2^\mathrm{L}},e^{-i\phi_1^\mathrm{L}}]$ and $T^\mathrm{R}_\phi=\mathrm{diag}[e^{i(\phi_2^\mathrm{L} + \phi_1^\mathrm{L}-\phi_3)}, e^{-i\phi_2^\mathrm{R}},e^{-i\phi_1^\mathrm{R}}]$ are
phase shifts before and after scattering, which are not independent,
and $\bar{T}$ is the unitary matrix 
       
\begin{equation}
\label{MatrixT}
\bar{T}=
\begin{pmatrix}
\cos{\alpha} e^{i\chi} & \frac{\sin{\alpha}}{\sqrt{2}}  & \frac{\sin{\alpha}}{\sqrt{2}}\\
\frac{\sin{\alpha}}{\sqrt{2}} & -\frac{1+ \cos{\alpha} e^{- i \chi}}{2} & \frac{1- \cos{\alpha} e^{- i \chi}}{2}\\
\frac{\sin{\alpha}}{\sqrt{2}} & \frac{1- \cos{\alpha} e^{- i \chi}}{2} &  -\frac{1+ \cos{\alpha} e^{- i \chi}}{2}
\end{pmatrix}.
\end{equation}
Here $\chi=\arccos[\{3 \cos^2(\alpha)-1\}/2\cos(\alpha)]$. 
The angle $\alpha$ defines the ratio of scattering probabilities between forward $P_\mathrm{f}$ and 
deflected $P_\mathrm{d}$ channels by $P_\mathrm{f}/P_\mathrm{d}=2 \cot^2(\alpha)$. 

The outgoing and incoming electronic waves at a node are connected by
 $\psi_\mathrm{out}=e^{-i \phi_\mathrm{E}} T  \psi_\mathrm{in}$, where $\psi_\mathrm{out}=(\psi_{i+1,j}^\mathrm{1}, \psi_{i,j-1}^\mathrm{2},\psi_{i,j}^\mathrm{3})$ and $\psi_\mathrm{in}=(\psi_{i,j-1}^\mathrm{1}, \psi_{i,j}^\mathrm{2},\psi_{i+1,j}^\mathrm{3})$.  Here $\phi_\mathrm{E}=\epsilon L /\hbar v_{||}$ is the dynamical phase accumulated 
by electrons while propagating between links. Bloch's theorem connects wave function amplitudes
in different cells by $\psi_{ij}=e^{i \vec{q}\vec{R}_{ij}} \bar{\psi}$, where 
$\bar{\psi}\equiv\{\bar{\psi}^1, \bar{\psi}^2,\bar{\psi}^3\}$ and $\vec{q}$ is the moir$\mathrm{\acute{e}}$ momentum.
The connection between input and output waves can be  
written as $[\lambda-U_\vec{q}]\bar{\psi}=0$,
and has a nontrivial solution only if $\lambda=e^{i ( \phi_\mathrm{E} - \phi_\mathrm{T})}$ is equal to 
one of eigenvalues of the matrix
\begin{widetext}
\begin{equation}
\label{MatrixU}
U_\vec{q}=\left(
	\begin{array}{ccc}
	\cos{\alpha} e^{\bm{i}(\chi+\phi_1^\mathrm{R}+\phi_2^\mathrm{R}+\phi_1^\mathrm{L}+\phi_2^\mathrm{L} - \vec{q}\vec{l}_1 - \vec{q}\vec{l}_2   )}  & \frac{\sin{\alpha}}{\sqrt{2}} e^{i (\phi_1^\mathrm{R}+ \phi_3 - \vec{q}\vec{l}_1 )}
		& \frac{\sin{\alpha}}{\sqrt{2}} e^{i (\phi_2^\mathrm{R}+\phi_3)} \\
	\frac{\sin{\alpha}}{\sqrt{2}}  e^{i (\phi_1^\mathrm{L}-\phi_3)} & -\frac{1+ \cos{\alpha} e^{- \bm{i} \chi}}{2} e^{i (\vec{q}\vec{l}_2 - \phi_2^\mathrm{R} - \phi_2^\mathrm{L})} &\frac{1- \cos{\alpha} e^{- \bm{i} \chi}}{2} e^{i (\vec{q} \vec{l}_1 +\vec{q} \vec{l}_2 - \phi_1^\mathrm{R}- \phi_2^\mathrm{L})} \\
	\frac{\sin{\alpha}}{\sqrt{2}}  e^{i (\phi_2^\mathrm{L}-\phi_3-\vec{q} \vec{l}_2)} & \frac{1- \cos{\alpha} e^{- \bm{i} \chi}}{2}  e^{-i (\phi_2^\mathrm{R}+\phi_1^\mathrm{L})} & -\frac{1+ \cos{\alpha} e^{- \bm{i} \chi}}{2} e^{i (\vec{q}\vec{l}_1 - \phi_1^\mathrm{R}-\phi_1^\mathrm{L})} \\
	\end{array}
	\right).
	\end{equation}
\end{widetext}
It follows that the electronic spectrum consists of groups of three bands $n=-1,0,1$ 
that repeat in energy with period $\epsilon_\mathrm{L}^{||}=2\pi\hbar v_{||}/L$  and 
have dispersion 
\begin{equation}
\label{BandStructure}
\epsilon^{nm}_\vec{q}=\epsilon_\mathrm{L}^{||}\;\left(\frac{\arg[\lambda^n_\vec{q}]}{2\pi}+\frac{\phi_\mathrm{T}}{2 \pi}+ m\right).
\end{equation}
Here $\lambda^n_\vec{q}$  are the eigenvalues of $U_\vec{q}$ and $m$ is an integer. The role of the phase $\phi_\mathrm{T}$ is just a rigid shifts of all bands in energy. Since the matrix $U_\vec{q}$ is also unitary 
$U_\vec{q}^+=U_\vec{q}^{-1}$ and $\det[U_\vec{q}]=1$, its eigenvalues satisfy  
\begin{equation}
\label{EigenvalueProblem}
\lambda_\vec{q}^3-\mathrm{tr}[U_\vec{q}]\lambda_\vec{q}^2+\mathrm{tr}[U_\vec{q}^+]\lambda_\vec{q}-1=0.
\end{equation}
The electronic spectrum therefore depends  only on
\begin{figure}[b]
	\label{Fig4}
	\vspace{-2 pt}
	\includegraphics[width=7.6 cm]{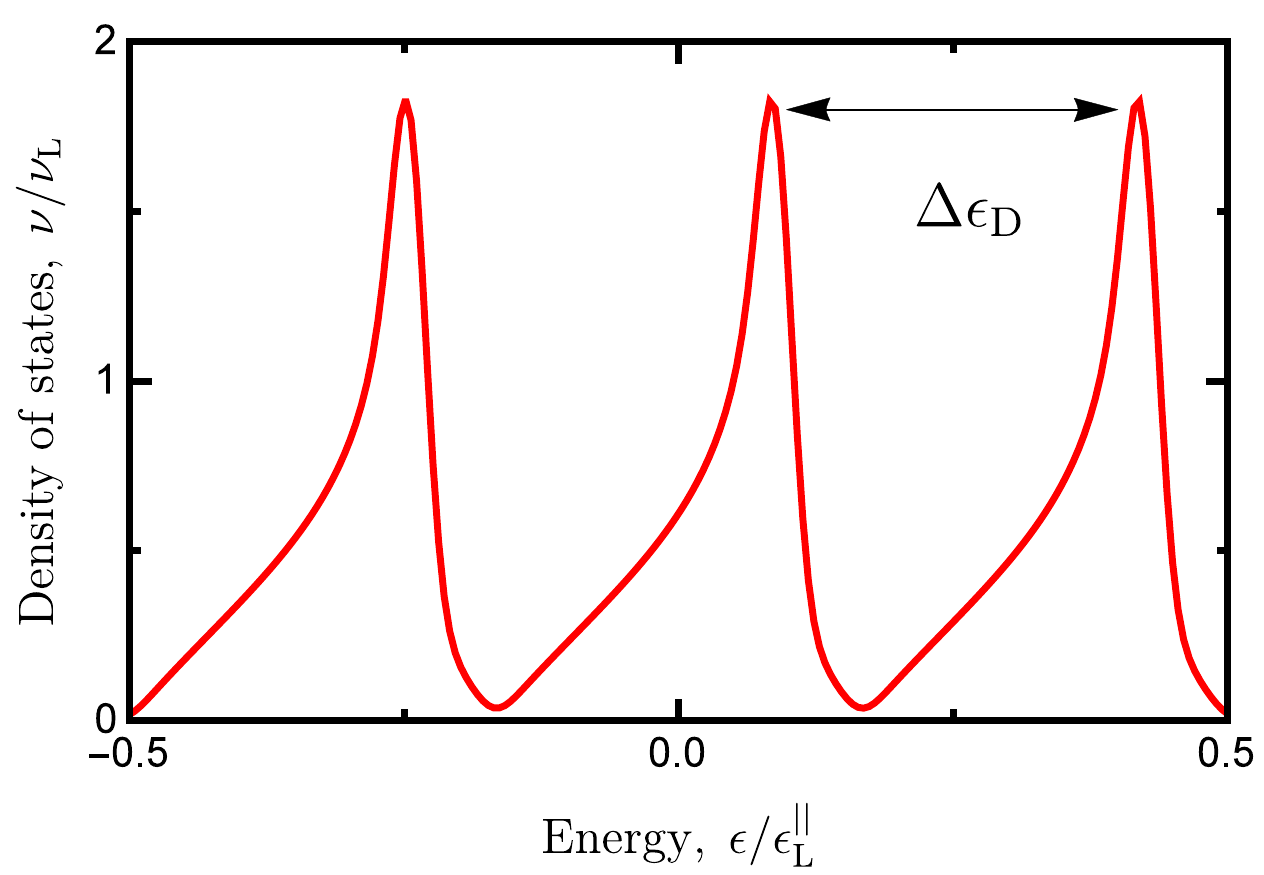}
	\vspace{-2 pt}
	\caption{The energy dependence of the density of states $\nu(\epsilon)$ per valley, spin and per Dirac point in the ring. It has three dips and three maxima separated from each other by $\Delta \epsilon_\mathrm{D}=\epsilon_\mathrm{L}^{||}/3$. The primer correspond to Dirac points, while the latter to saddle points of the moir$\mathrm{\acute{e}}$ pattern band structure presented in Fig.~1. The corresponding scale for the density of states is $\nu_\mathrm{L}=\sqrt{3}\pi/ \epsilon_\mathrm{L}^{||}L^2 $ .}
\end{figure}
\begin{equation}
\label{TrU}
\begin{split}
\mathrm{tr}[U_\vec{q}]=\cos(\alpha) e^{\bm{i} \chi} e^{i (\Phi_1+\Phi_2-\vec{q} \vec{l}_1 -\vec{q} \vec{l}_2 )}\\
-\frac{1}{2}\left[1+\cos(\alpha) e^{- i \chi}\right] \left[e^{i (\vec{q} \vec{l}_1 - \Phi_1)}+e^{i (\vec{q} \vec{l}_2 - \Phi_2)}\right].
\end{split}
\end{equation}  
Here we have introduced phases $\Phi_1=\phi_1^\mathrm{L}+\phi_1^\mathrm{R}$, $\Phi_2=\phi_2^\mathrm{L}+\phi_2^\mathrm{R}$. These phases $\Phi_1$ and $\Phi_2$ can be eliminated by the shift of the momentum space origin, and therefore do not influence the density of states of the network and electronic transport through it. The latter remarkably depend only on $\alpha$, which in turn characterizes the distribution of scattering probability between 
forward and deflected channels. It has been numerically 
shown~\cite{QiaoJungPRL2014} that, contrary to classical intuition, 
because nearby paths
have larger wavefunction overlap with the incoming electron,
deflection is the more likely outcome.
For presentation of results we chose $\alpha=1.1$ corresponding to 
$P_\mathrm{f}\approx 0.2$ and $P_\mathrm{d}\approx 0.4$.

The first Brillouin zone of the network has a hexagonal shape and is illustrated
in Fig.3-c where we also illustrate an equivalent rhombic primitive cell. 
The spectrum has the mirror symmetry across the $\mathrm{K}\mathrm{K}'$ line since
$\mathrm{tr}[U_{q_\mathrm{M}-q_x,q_y}]=\mathrm{tr}[U_{q_\mathrm{M}+q_x,q_y}]$, where 
$\vec{q}_\mathrm{M}=2\pi \vec{e}_\mathrm{x}/\sqrt{3}L$ is the position of the $\mathrm{M}$-point in the 
Brillouin zone. For presentation of results we have chosen $\phi_\mathrm{T}=\Phi_1=\Phi_2=(\pi-2\arcsin[3\sin\alpha/2\sqrt{2}])/3$ that ensures the discrete rotational symmetry of the network band structure with respect to $120^\circ$ around the $\Gamma$-point. 
A single period $\epsilon^{n0}_\vec{q}$ of the repeating band structure is plotted in 
the half of the rhombic Brillouin zone in Fig.~1, where we see that it is gapless because of Dirac band touching points situated in $\Gamma$, $\hbox{K}$, and $\hbox{K}'$ high symmetry points. Their positions are independent on $\alpha$ and they are separated by momentum $\Delta k_\mathrm{D}=4\pi/3L$ and energy $\Delta \epsilon_\mathrm{D}=\epsilon_\mathrm{L}^{||}/3$. The density of states of the network is presented in Fig.~4
and is periodic with period $\Delta \epsilon_\mathrm{D}$. It is three time smaller than the period of the network band structure $\epsilon_\mathrm{L}^{||}$, that reflects the symmetry between three links in an elementary cell of the model. The single period contains one zero at the Dirac point, and one 
saddle-point logarithmic divergence. The latter reflects the van Hove singularity due to the presence of saddle points in the network band
structure, which are clearly visible in Fig. 1.  

In recent experiments~\cite{2018arXiv180202999H} the small twist-angle $\theta=0.245^{\circ}$ has been applied between layers and has resulted in moir$\mathrm{\acute{e}}$ patterns with period $L\approx58\;\mathrm{nm}$. The resulting energy scale of the pattern $\epsilon_\mathrm{L}=2\pi \hbar v/L\approx 72 \; \hbox{meV}$ is comparable with the induced gap $\epsilon_\mathrm{g}\approx 60\; \hbox{meV}$. While the phenomenological network model is still reasonable at energies $\epsilon\ll \epsilon_\mathrm{g}$, the expressions for  $v_\mathrm{||}$ and $\Delta \epsilon_\mathrm{D}$  do not directly apply. Our model predicts the periodic set of features in the density of states, whereas only one feature within the gap has been observed~\cite{2018arXiv180202999H}.  For the gap $\epsilon_\mathrm{g} \approx 250 \; \hbox{meV}$ achievable in bilayer graphene~\cite{BGgap1,BGgap2}, our model is well applicable in much wider range of energies. Using the hybridization energy  $w=400\; \hbox{meV}$ we get that the velocity of helical states $v_{||}=1.6\; 10^6 \; \mathrm{m}/\mathrm{s}$ is larger than the velocity of electrons in graphene $v=10^6 \; \mathrm{m}/\mathrm{s}$. The period of the network is equal to $\epsilon_\mathrm{L}^{||}\approx 115\; \hbox{meV}$ and the the period of density of states $\Delta \epsilon_\mathrm{D}\approx 38 \; \hbox{meV}$. It is much smaller than the gap $\epsilon_\mathrm{g}$ and we expect a set of features due to van Hove singularities of network spectrum to be well resolved in experiments. Alternatively, the condition $\epsilon_\mathrm{L}^{||}\ll\epsilon_\mathrm{g}$ can be achieved at smaller twist angles $\theta$. 

To conclude, we have introduced a new phenomenological network model which captures the electronic structure of twisted bilayer graphene in the energy range 
below the $\mathrm{AB}$ and $\mathrm{BA}$ gaps where only topologically confined domain wall states are present. Motivated by the recent observation of the domain wall network in STM experiments~\cite{2018arXiv180202999H} we have focused on its band structure and density of states. Very recently signatures of the network formation have been found in magneto-transport experiments~\cite{2018arXiv180207317R}. Whereas our model predicts anisotropic transport properties that are approximately periodic in carrier density, the magneto-transport theory is postponed for future work. 

\emph{Acknowledgment}.  This material is based upon work supported by the Department of Energy under Grant No DE-FG02-ER45118 and by the Welch Foundation under Grant No. F1473. 

\bibliography{MoireNetworkBIB}

\begin{thebibliography}{53}
\bibitem [{Not({\natexlab{a}})}]{NoteShift}%
  \BibitemOpen
  \href@noop {} {} ({\natexlab{a}}),\ \bibinfo {note} {a geometry of the
  graphene bilayer needs to be characterized not only by a twist $\theta$, but
  also by a relative translation vector $\vec{d}$. As it has been shown
  previously~\cite{MacDonaldBistritzer} the latter only shifts the
  moir$\mathrm{\acute{e}}$ pattern, while its electronic band structure is
  independent on $\vec{d}$.}\BibitemShut {Stop}%
\bibitem [{Not({\natexlab{b}})}]{NoteTd}%
  \BibitemOpen
  \href@noop {} {} ({\natexlab{b}}),\ \bibinfo {note} {the isotropic
  contribution $t_\mathrm{S}$ is important only in the vicinity of domain wall
  intersections. In our phenomenological network model the scattering matrix
  for nodes~$T$ is parametrized in a phenomenological way and details of the
  intersections profile are off importance.}\BibitemShut {Stop}%
\end{thebibliography}

\end{document}